\documentclass[aps,pra,twocolumn,superscriptaddress,amsmath,amssymb,floatfix,showpacs]{revtex4}

\usepackage{color}
\usepackage{graphicx}
\usepackage{dcolumn}
\usepackage{bm}
\usepackage[colorlinks=true,linkcolor=blue]{hyperref}%

\newcommand{\sslash}{\mathbin{/\mkern-4mu/}}

\graphicspath{{fig-ps/}}

\hypersetup{
 citecolor=blue,
 linkcolor=magenta,
 urlcolor=magenta}

\begin{document}

\preprint{AIP/123-QED}

\title[Scattering and leapfrogging of vortex rings in a superfluid]
{Scattering and leapfrogging of vortex rings in a superfluid}

\author{R. M. Caplan}
\affiliation{Predictive Science Inc.\footnote{{\tt URL:} http:$\sslash$www.predsci.com/},
9990 Mesa Rim Road Suite 170,
San Diego, California 92121, USA}
\author{J. D. Talley}
\author{R. Carretero-Gonz\'alez}
\affiliation{
Nonlinear Dynamical Systems Group\footnote{{\tt URL:} http:$\sslash$nlds.sdsu.edu/},
Computational Science Research Center\footnote{{\tt URL:} http:$\sslash$www.csrc.sdsu.edu/},
and Department of Mathematics and Statistics,
San Diego State University, San Diego, California 92182-7720, USA}
\author{P.G. Kevrekidis}
\affiliation{%
Department of Mathematics and Statistics,
University of Massachusetts,
Amherst, MA 01003-4515, USA
}

\date{\today}

\begin{abstract}
The dynamics of vortex ring pairs in 
the homogeneous nonlinear Schr\"odinger equation
is studied.  The generation of numerically-exact solutions of 
traveling
vortex rings is described and their translational velocity compared to revised analytic approximations.
The scattering behavior of co-axial vortex rings with opposite charge 
undergoing collision is numerically investigated 
for different scattering angles yielding a surprisingly simple result for 
its dependence as a function of the initial vortex ring parameters.  We also study the leapfrogging behavior of co-axial rings with equal charge and compare 
it with the dynamics stemming 
from a modified version of the 
reduced equations of motion from a classical fluid model derived using the 
Biot-Savart law.
\end{abstract}

\maketitle

\section{Introduction}
\label{s:intro}
One of the most widespread and interesting models for studying
the emergence, dynamics and interactions of coherent
structures is the nonlinear Schr{\"o}dinger equation (NLSE)
\cite{sulem99}.
The NLSE is a paradigm for the evolution of
coherent structures since it is a universal model
describing the evolution of complex field envelopes in nonlinear
dispersive media \cite{dodd83}.
This universality stems from the fact that the NLSE is the prototypical,
lowest-order, nonlinear partial differential equation (PDE) that describes
the dynamics of modulated envelope waves in a nonlinear medium
\cite{debnath05}.
As such, 
it appears
in a wide variety of physical contexts,
ranging from optics \cite{hasegawa90,abdullaev93,hasegawa95,kivshar03}
to fluid dynamics and plasma physics \cite{infeld90}
to matter waves \cite{kevrekidis08,carretero08b}, while it has
also attracted much interest
mathematically~\cite{sulem99,ablowitz04,bourgain99,ablowitz81,zakharov84,newell85}.

The general form of the NLSE (with the lowest order nonlinearity)
can be written, in non-dimensional units, as
\begin{equation}
\label{nlse}
i\frac{\partial \Psi}{\partial t}+a\nabla^2\Psi
+s|\Psi|^2\Psi=0
\end{equation}
where $a$ is a parameter and $s$
represents the strength of the nonlinear interaction
and determines whether the NLSE is attractive ($s>0$) or repulsive ($s<0$).
The NLSE is a model for the evolution of the mean-field
wavefunction in Bose-Einstein condensates (BECs) at low
temperatures \cite{Dalfovo99,Pethick2002}, 
namely a superfluid.
In the BEC setting, the
NLSE also contains an external trapping potential
term that is
often spatially parabolic or periodic in nature and
stems from the magnetic or optical confinement of the atoms.

The NLSE allows for the prediction and description of a wide range
of (nonlinear) excitations depending on the sign of
the nonlinearity and the dimensionality of
the system, the latter of which that can be tuned by the ratio of trapping strengths
along the different spatial directions~\cite{kevrekidis08}.
In this manuscript we are interested in the fully three-dimensional
regime (3D) with repulsive nonlinearity ($s<0$) whose basic
nonlinear excitations correspond to vortex lines and vortex rings.
A vortex line is the 3D extension of a two-dimensional (2D) vortex
by (infinitely and homogeneously) extending the solution into the
axis perpendicular to the vortex plane.
Vortex lines might be rendered finite in length if the background where they
live is bounded by the externally confining potential.
In that case, vortex lines become vorticity ``tubes'' that are straight across
the background or bent in {\tt U} and {\tt S} shapes, depending on the aspect
ratio of the background \cite{VR:USshapedVLs1,VR:USshapedVLs2}.
If a vortex line is bent enough to close on itself, or if two
vortex lines are close enough to each other, they can produce
a vortex ring \cite{VR:CrowInstab}.
Vortex rings are 3D structures whose core is a closed loop with vorticity
around it \cite{donnelly} (i.e., a vortex line that loops back into itself).
Vortex rings have been observed experimentally in superfluid helium \cite{Rayfield64, Gamota73} as well as in the context of BECs in the decay of dark solitons in two-component BECs \cite{Anderson01}, direct density engineering \cite{Shomroni09} (see also the complementary theoretical proposals
of Refs.~\onlinecite{Ruostekoski05,Ruostekoski01}), 
and in the evolution of colliding symmetric defects \cite{Ginsberg05}.  
They have also been argued to be responsible for unusual experimental
collisional
outcomes of structures that may appear as dark solitons in cigar-shaped
traps~\cite{sengstock}.
Numerical studies of experimentally feasible vortex ring generation in BECs have also been explored in the context of flow past an obstacle \cite{Jackson99,Rodrigues09}, Bloch oscillations in an optical trap \cite{Scott04}, collapse of bubbles \cite{Berloff04a}, instability of 2D rarefaction pulses \cite{Berloff02}, flow past a positive ion \cite{Berloff00, Berloff00a} or an electron bubble \cite{Berloff01}, crow instability of two vortex pairs \cite{Berloff01a} and
 collisions of multiple BECs \cite{Carretero08,Carretero08a}.
Vortex rings have an intrinsic velocity \cite{Roberts71}, which can be overcome by counteracting the velocity with a trapping potential \cite{VR-BEC-STRUCTGOOD}, adding Kelvin mode perturbations \cite{maggioni10,Helm11,Helm10}, or by placing the vortex ring in a co-traveling frame (see Sec.~\ref{s:1VR}).

In the present manuscript, motivated by the above abundant interest in this
theme both from a theoretical perspective and from that of
ongoing experimental efforts in BECs, we are interested in the dynamical
evolution of vortex rings and in particular in their interactions.
In an effort to exclusively capture the vortex ring dynamics
and not the influence of the external potential (the latter is 
a natural subject for future work), we assume
the absence of any trapping potential 
and thus consider
the background supporting the vortex rings as homogeneous.
On this homogeneous background we revisit the intrinsic
velocity of a single vortex ring that serves as the
self-interacting term that will be supplemented by
the interaction terms between different vortex rings.
The main goal of our work is to describe vortex ring interactions
and scattering collisions in a reduced (i.e., ``effective'') manner.
The motivation for this approach stems from the fact that
full numerical simulations in the 3D NLSE 
can be very time consuming especially 
when examining scenarios involving multiple vortex
rings in large domains.
The first dynamical scenario that we study corresponds to the
scattering of colliding vortex rings of opposite charge.
Using a large set of full 3D numerical simulations (based on
efficient GPU codes, see below), we are able to distill
a very simple, phenomenological, rule describing the
scattering angle of vortex rings as a function of the
vortex ring radii and the initial collisional offset.
Then, based on the interaction of vortex filaments from
classical fluids, we revisit a reduced, ordinary
differential equation, model for the interaction of
co-axial vortex rings based on the Biot-Savart law,
adapting it to the superfluid (BEC) case.
The ensuing reduced dynamics for the leapfrogging of vortex ring
pairs is then favorably compared to full 3D simulations.

Our manuscript is structured as follows. In Sec.~\ref{s:1VR} we describe a procedure to generate numerically-exact traveling vortex ring solutions in a homogeneous background, and test the resulting ring's translational velocity against 
suitably revised known analytic approximations.
Section~\ref{s:scattering} is devoted to describing
scattering scenarios from the collisions of vortex rings, where
we derive a very simple phenomenological relationship for
the scattering angle as a function of the initial
offset distance and radii of the colliding rings.
In Sec.~\ref{s:leapfrog} we use effective equations
of motion for the interaction of co-axial vortex
rings to describe the leapfrogging behavior of
two vortex rings and compare the results with direct simulations of the NLSE.
Finally, Sec.~\ref{s:conclu} summarizes our
results and prompts a few avenues for
future exploration.

Before we embark on the relevant analysis, we provide some details
on our computational methods which pertain to all the numerical simulations
given below.
For all 3D NLSE simulations, we use high-order explicit finite-difference schemes on a uniform grid with cell-spacing $h=\Delta x=\Delta y=\Delta z$ and constant time-step $k=\Delta t$.  Time-stepping is accomplished with the standard 4th-order Runge-Kutta method (RK4), while spatial differencing is performed with the 4th-order 2-step high-order compact scheme described in Ref.~\onlinecite{Caplan13}.  The time-step $k$ is set based on the stability bounds of the overall scheme as discussed in Ref.~\onlinecite{Caplan13a}. Since the stability forces the time-step to be proportional to $h^2$, the overall error of the scheme is $O(h^4)$.  Due to the constant density background of the problem, we utilize a recently developed modulus-squared Dirichlet boundary condition \cite{Caplan14}.  The simulations are computed using the NLSEmagic code package \cite{NLSEmagic} \footnote{{\tt URL:} http:$\sslash$www.nlsemagic.com} which contains algorithms written in C and CUDA, and are primarily run on NVIDIA GeForce GTX 580 and GeForce GT 650M GPU cards.

\section{Generation of a numerically-exact non-stationary vortex ring}
\label{s:1VR}
In order to achieve accurate results for simulating the interactions of multiple vortex rings in a homogeneous background, it is necessary to first be able to generate numerically ``exact'' solutions of the individual rings.  Since vortex rings have an intrinsic transverse velocity due to their topological structure, standard steady-state methods (such as imaginary time integration \cite{VR-BEC-STRUCTGOOD} and nonlinear equation solvers) for obtaining solutions of coherent structures cannot be directly applied, unless the rings are in a configuration where they are at steady-state (such as in a magnetic trap \cite{VR-BEC-STRUCTGOOD}).  In order to be able to generate rings in a homogeneous background, we apply a co-moving background velocity to render the vortex ring stationary.

In order to apply the back-flow, we need a very accurate value of the vortex ring's transverse velocity.  The transverse velocity of a single vortex ring in the NLSE has been studied analytically \cite{Amit66,Fetter66,Roberts71} and the results were shown to be consistent with numerical simulations of the NLSE \cite{Koplik96,Helm10}.  For the NLSE in the form of Eq.~(\ref{nlse}), with $s<0$ and $V=0$, an asymptotically approximate velocity for the ring is given by \cite{Roberts71}
\begin{equation}
\label{vrvel}
c \approx -\frac{am}{d}\left(\ln \frac{8d}{r_c} + L_0(m) - 1\right),
\end{equation}
where $d$ is the ring's radius, $m$ is its the charge, 
and $r_c$ is the vortex ring's core radius defined as
\begin{equation}
r_c=|m| \xi,
\end{equation}
where $\xi$ is the healing length given by
\begin{equation}
\xi=\sqrt{-\frac{\Omega}{a}}.
\end{equation}
$\Omega$ plays the role of the frequency
and is tantamount to the system's chemical potential.
The value $L_0(m)$ is referred to as the vortex core parameter and is defined as the convergent part of the energy-per-unit-length of a vortex line in the NLSE
\cite{Roberts71}.  While  $L_0(m)$ depends on the vortex ring charge $m$, it is independent of the NLSE parameters.  In order to determine the velocity of the vortex ring as precisely as possible, $L_0(m)$ must be computed accurately.
Since there are various values of $L_0(m)$ reported in the literature
(cf. values given in Refs.~\onlinecite{VLINE-L058,Roberts71,VR-ROBERTS-BOOK}),
we briefly show in the Appendix how $L_0(m)$ is numerically computed and
give its values for $m\in[1,10]$.
Since we are only focusing on vortex rings of charge $|m|=1$, (indeed, stable
higher-charge vortex rings are not known to exist \cite{VLINE-L0-61}),
we use the value $L_0(1)\approx 0.380868$.

\begin{figure}[htb]
\centering
\includegraphics[width=7.5cm]{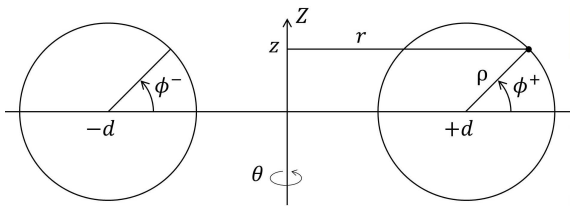}
\caption{
Cylindrical coordinates used to describe the initial condition for
a vortex ring.
}
\label{f:cylindrical}
\end{figure}

The initial condition for a single vortex ring solution in cylindrical
coordinates (see Fig.~\ref{f:cylindrical}) with radius $d$ centered
at $(r,z)=(0,0)$ can be described by
$$
\Psi_0 = \Psi(r,z,\theta,0) = g(r,z)\,\exp[im(\phi^{-}(r,z)-\phi^{+}(r,z))],
$$
where
$$
\phi^{\pm}(r,z) = \mbox{arctan}\left(\frac{z}{r\mp d}\right),
$$
and the function $g(r,z)$ is an axisymmetric 2D profile which may contain additional phase information of the solution.
This 2D profile can be approximated by the modulus of the radial
solution of a 2D NLSE vortex in the $(r,\phi^{+})$ plane and is described by
$$
g(r,z)=f(\rho(r,z)),
$$
where $\rho(r,z) = \sqrt{(r-d)^2 + z^2}$ and $f(\rho)$ is the numerical solution to the radial steady-state equation
$$
-\left(\Omega + \frac{am^2}{\rho^2}\right)f(\rho) + a\left(\frac{1}{\rho}\frac{df}{d\rho} + \frac{d^2f}{d\rho^2}\right) + s\,f^3(\rho)=0,
$$
which is derived from inserting the form of a vortex solution
$\Psi=f(\rho)\,\exp[i(m\phi^{+}+\Omega t)]$
into the 2D NLSE in polar coordinates.

In order to obtain a numerically-exact vortex ring initial condition, we would like to solve a steady-state equation for $\Psi_0$ using the above approximation as an initial seed.  This can be accomplished by noting that a steady-state solution to the NLSE of Eq.~(\ref{nlse}) in cylindrical coordinates in a co-moving frame in the $z$-direction with velocity $c$ is given by \cite{RMC-DISS}
$$
\Psi = U(r,Z,\theta)\,\exp\left[i\left( \frac{c}{2a}z + \left[\Omega - \frac{c^2}{4a}\right]t \right)\right],
$$
where $Z = z-ct$ and $U(r,Z,\theta)$ solves the time-in\-de\-pen\-dent NLSE
\begin{equation}
\label{comoveSSnlse}
-\Omega U + a\left(\frac{1}{r}\frac{\partial U}{\partial r} + \frac{\partial^2 U}{\partial r^2} + \frac{\partial^2 U}{\partial z^2}\right) + s|U|^2U = 0.
\end{equation}
Therefore, by imposing a counter-flow with a velocity equal and
opposite to that of the intrinsic velocity of the vortex ring by
the following Galilean boost:
$$
U_0(r,z) = \Psi_0\,\exp\left(-i\frac{c}{2a}z\right),
$$
where $c$ is the analytical approximation of the vortex
ring velocity given by Eq.~(\ref{vrvel}),
we can then solve Eq.~(\ref{comoveSSnlse}) for $U$.
We use the nonlinear Newton-Krylov solver {\tt nsoli} \cite{OPT-NSOLI-BOOK}
to find the solution, using a central-difference discretization of the spatial
derivatives, along with the modulus-squared Dirichlet boundary conditions
mentioned in Sec.~\ref{s:intro}. The initial condition of the vortex
ring solution is then found by removing the added counter-flow velocity:
$$
\Psi(r,z,\theta,0) = U\,\exp\left(i\frac{c}{2a}z\right).
$$

\begin{figure}[htb]
\centering
\includegraphics[width=7cm]{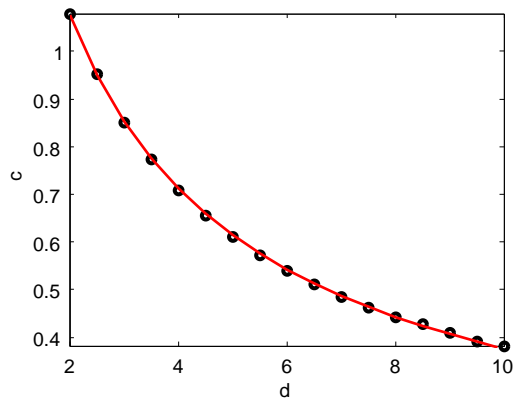}
\caption{%
(Color online)
Velocity for a vortex rings of charge $m=1$ as a function of
its radius. The dots are the velocities computed from direct 3D
simulations averaged over a $t\in[0,50]$, while the line is the
predicted velocity of Eq.~(\ref{vrvel}). The NLSE parameters used
are $a=1$, $s=-1$, and $\Omega=-1$, while the numerical parameters
are $h=0.5$ (grid spacing) and $k=0.025$ (time step).\label{f:VRvel}}
\end{figure}

\begin{figure*}[htb]
\centering
\includegraphics[width=\linewidth]{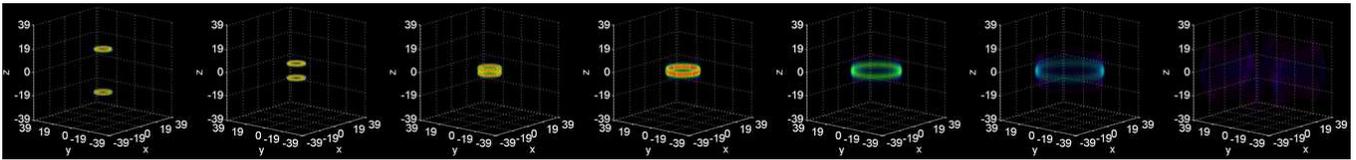}
\caption{
(Color online)
Example of a head-on collision between vortex rings of charge $m=1$
and $m=-1$.  The rings have a radius of $d=6$ and are shown at times
$t=0,22,33,39,45,51,59,68$. The images depict an inverse
transparency volumetric rendering of the density $|\Psi|^2$
such that $|\Psi|^2=1$ is fully transparent and $|\Psi|^2=0$ is fully opaque.
The NLSE parameters in the
simulation are $a=1$, $s=-1$, and $\Omega=-1$, while the numerical
parameters are $h=0.8$ and $k=0.065$.\label{f:headoncol}}
\end{figure*}

The numerically exact vortex ring solutions can be used to compare the analytical approximations of the vortex ring's transverse velocity Eq.~(\ref{vrvel}) to direct simulations of the NLSE.  In order to track the vortex rings during the simulations, a 2D $(y,z)$ cut of the computational grid is extracted, and a center-of-mass calculation of the modulus of the vorticity in the region of the ring's intersection points in the cut are tracked.  In Fig.~\ref{f:VRvel} we show the results of integrating vortex rings until $t=50$ for radii $d \in [2,10]$ and comparing their tracked velocity to Eq.~(\ref{vrvel}).
It is seen that the direct velocity results match the analytical approximation very accurately; the disagreement is never more than $1.5\%$ in our studies.  This illustrates that values obtained from the asymptotic velocity equation, Eq.~(\ref{vrvel}), match direct integration even for rings with relatively small radii.  We note that although the velocity $c$ is used to generate the vortex rings initially, the fact that the rings maintain the velocity over a long integration time with no noticeable noise or counter-flow in the solution confirms that the simulations demonstrate the validity of Eq.~(\ref{vrvel}).

We note here that the transverse velocity discussed in this section is that of an {\em unperturbed} vortex ring.
It is possible to perturb a vortex ring (or line) to produce oscillations along
the ring called Kelvin modes (or Kelvons) \cite{fetter04,simula08,horng06,chevy03}, one natural method being the merger and scattering of two vortex rings (discussed in the next section).  These Kelvin modes not only have their own dynamics, interactions
and decay properties~\cite{garcia01,Barenghi:arXiv13}, but they can also self-interact within a vortex ring, resulting in a reduction or even reversal of its transverse velocity \cite{maggioni10,Helm11,Helm10}.

\section{Scattering vortex rings}
\label{s:scattering}
We now present a quantitative analysis of the scattering of two unit-charge ($|m|=1$) vortex rings in the NLSE.
An initial qualitative study of the scattering of such rings was performed in Ref.~\onlinecite{Koplik96}, where the authors focused on head-on collisions of the rings at different angled orientations.  In contrast, we focus here exclusively on offset co-planar collisions, not studied in Ref.~\onlinecite{Koplik96}.

As shown in Ref.~\onlinecite{Koplik96}, two colliding vortex rings with zero offset (axisymmetric) expand and annihilate each other's topological charge resulting in an axisymmetric decaying rarefaction pulse.  An example of such a case is shown in Fig.~\ref{f:headoncol} where two vortex rings of radius $d=6$ (one with charge $m=1$ and one with $m=-1$) are positioned a distance of $6d$ apart from each other in the $z$-direction and allowed to collide.
We study here the scattering that results when the colliding vortex rings are 
offset from each other in the planar direction.
Our numerical experiments show that, typically, the two rings merge and then separate into two new vortex rings moving away from each other at an angled trajectory, each exhibiting large quadrupole Kelvin 
oscillations.

\begin{figure}[htb]
\centering
\includegraphics[width=7.5cm]{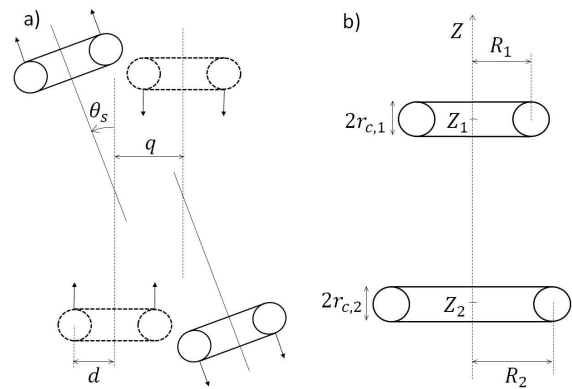}
\caption{
(a) Schematic of a scattering scenario between two co-planar vortex rings
of initial radius $d$ offset by a distance $q$ (impact parameter).
The initial, pre-scattering, configuration is depicted with dashed lines
while the post-scattering configuration is depicted with solid lines.
The scattering angle is defined as the, signed, angle between incoming
and outgoing axes as shown in the schematic.
(b) Schematic of the variables to describe the reduced dynamics for
the leapfrogging between two co-axial vortex rings located at $Z_1$ and
$Z_2$ with respective radii $R_1$ and $R_2$.
}
\label{f:VR_Drawing}
\end{figure}

\begin{figure*}[htb]
\centering
\includegraphics[width=\linewidth]{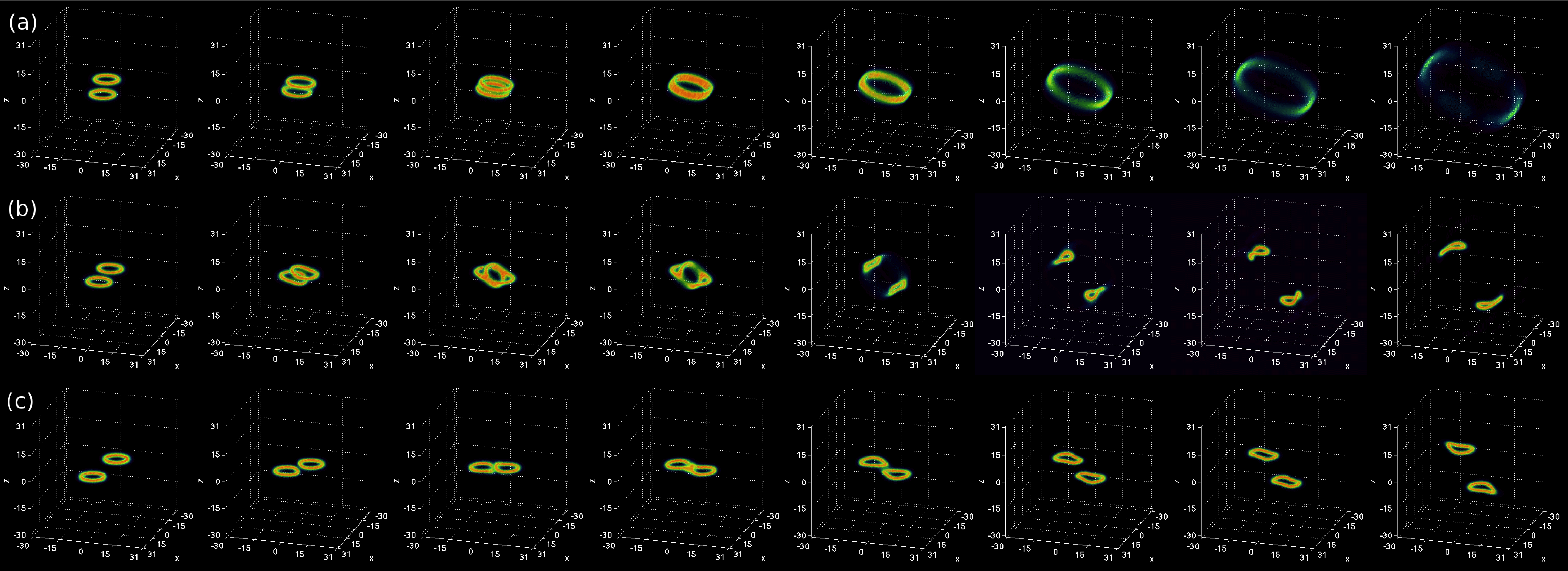}
\caption{%
(Color online)
Qualitatively distinct scattering scenarios for two colliding,
initially offset,
planar vortex rings corresponding to the three regions
depicted in Fig.~\ref{f:scatangvsos}.
(a) Annihilation scenario (with initial offset
of $q/d=0.35$) at times $t\approx4,8,12,16,19,23,27,32$.
(b) Merging/splitting collision (with initial offset of $q/d=1.0$)
at times $t\approx4,10,15,16,20,26,32,37$.
(c) Non-merging interactive fly-by (with initial offset of $q/d=2.2$)
at times $t\approx1,6,9,12,16,20,24,31$.
The NLSE and numerical parameters are the same as in Fig.~\ref{f:headoncol}.}
\label{f:scatqual}
\end{figure*}

\begin{figure}[htb]
\centering
\includegraphics[width=8cm]{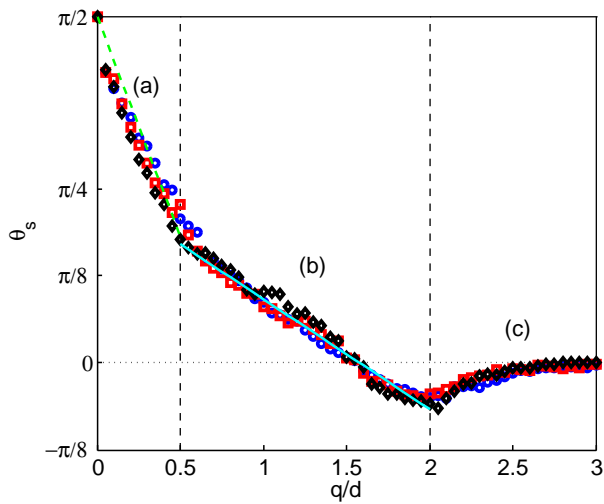}
\caption{
(Color online)
Scattering angle versus the normalized offset distance $q/d$ for
two colliding co-planar unitary-charge vortex rings.
The different regions correspond to the three qualitatively
scenarios depicted in Fig.~\ref{f:scatqual}.
The (blue) circles are the results for vortex rings of radius $d=6$,
the (red) squares for $d=8$ and the (black) diamonds for $d=10$.
The (green) dashed line is the approximation of Eq.~(\ref{scatpulse})
and the (cyan) solid line is the approximation of Eq.~(\ref{scatscat}).
The NLSE and numerical parameters used in the simulations are the same
as in Fig.~\ref{f:headoncol}.
\label{f:scatangvsos}}
\end{figure}

To quantify the relationship between offset distance and scattering angle, the vortex rings are positioned at a distance of $6d$ apart in the $z$-direction, and in the $y$-direction they are placed with a planar offset distance (impact parameter) defined as $q$ (see Fig.~\ref{f:VR_Drawing}(a)).  The scattering angle $\theta_s$ is defined as the angle away from vertical so that if the vortex rings travel in a straight trajectory, the angle is $0$, while if they scatter at right-angles to each other, the angle is $\pi/2$ (see Fig.~\ref{f:VR_Drawing}(a)).
Multiple simulations of colliding rings were performed with planar offsets ranging from $q/d=0$ to $q/d=3$ in increments of $0.05$ for vortex rings of radius $d=6,8,10$.
We found three different qualitative scenarios as depicted
and described in Fig.~\ref{f:scatqual}.
The center of the vortex rings in the $(y,z)$ plane are tracked and the scattering angle recorded.  The results are shown in Fig.~\ref{f:scatangvsos}.
It is clear from the results that there are 
three distinct sections in the relationship between scattering angle and offset, which correspond to qualitatively distinct topological outcomes as follows:

\begin{itemize}
\item[(a)]
For offsets with $q/d\in[0,1/2)$ the two rings annihilate each other's topological charge, sending out a rarefaction pulse,
see Fig.~\ref{f:scatqual}(a), similar to the cases of head-on collision shown in Fig.~\ref{f:headoncol}.  However in this case, as the rings approach each other, they rotate causing the collisions to occur at the scattering angle.  Another distinction of these collisions is that the resulting rarefaction pulse is not axi-symmetric as in the
head-on collision case, but more concentrated in the scattering direction.
\item[(b)]
When the offset has $q/d\in(1/2,2)$, the two rings undergo merging
immediately followed by splitting into internally excited rings with
quadrupole oscillations, (see Fig.~\ref{f:scatqual}(b)).  Interestingly,
when the offset is set to $q/d=3/2$ or larger, the scattering angle
becomes negative, but the qualitative behavior remains the same.
\item[(c)]
For larger offset such that $q/d>2$, the two rings no longer merge and
separate into new rings, but are merely perturbed by each other in a
fly-by scenario causing their trajectories to travel at a
small negative angle from incidence
(i.e., get weakly deflected ``inward''), 
see Fig.~\ref{f:scatqual}(c).  
As the offset increases,
this interaction decreases until approximately $q/d=3$ where the rings
essentially ignore each other's presence and travel unperturbed.
%
%
\end{itemize}

Noting that the scattering angle as a function of offset in Fig.~\ref{f:scatangvsos} appears linear in the first two qualitatively distinct intervals, simple phenomenological approximations of the relationships can be formulated.  For the annihilation section [$q/d\in(0,1/2)$], the angle of the rarefaction pulse's trajectory can be approximated by
\begin{equation}
\label{scatpulse}
\theta_s \approx -2\frac{q}{d} + \frac{\pi}{2},
\end{equation}
while for the scattering rings [$q/d\in(1/2,2)$], the angle of the resulting rings can be approximated as
\begin{equation}
\label{scatscat}
\theta_s \approx -\frac{1}{2}\frac{q}{d} + \frac{\pi}{4}.
\end{equation}
These approximations are shown along with the simulation data in Fig.~\ref{f:scatangvsos}.  It is interesting to note that, despite the complex dynamics involved in the scattering of vortex rings, we are able to formulate a very simple, phenomenological scattering rule for co-planar collisions.

It is interesting to contrast the 3D scenario of co-planar vortex ring
scattering to its 2D counterpart. If one focuses on the dynamics
restricted on the plane defined by the vortex ring axes, each 
vortex ring induces a vortex pair in 2D.
The case of scattering between 2D (point-) vortex pairs has been 
studied in some
detail~\cite{Aref-PTRSA-88,Price-PF-93,Aref-PF-08}. The most
notable differences between the ensuing dynamics for 2D 
vortex pairs and that of 3D vortex rings scattering is that the dynamics in 2D
is apparently richer due to the fact that truly 2D vortices are
more ``free'' to move (respecting the vortex-vortex interactions)
while the vortex pairs produced by the cut of the vortex ring in the
plane are more tightly linked to each other through the ring.
This extra freedom for 2D vortices allows for a rich
variety of dynamical evolution scenarios. For instance, depending of
the initial configuration (initial positions, orientations
and vortex pair internal distances) the collision dynamics 
of two vortex pairs can display:
(a) transient bound vortex pairs where the two vortex pairs
interact by orbiting around each other and exchange partners
for a few periods until scattering away from each other as
two distinct vortex pairs,
(b) transient interactions between the four vortices that are 
chaotic and lead to the eventual expulsion of two distinct vortex 
pairs resulting in chaotic scattering,
or
(c) transient bound states where three of the four vortices lock into a 
translational and precessional motion, while the fourth one moves separately, 
before the two pairs again scatter away, separating indefinitely.

\begin{figure}[htbp]
\centering
\includegraphics[width=\linewidth]{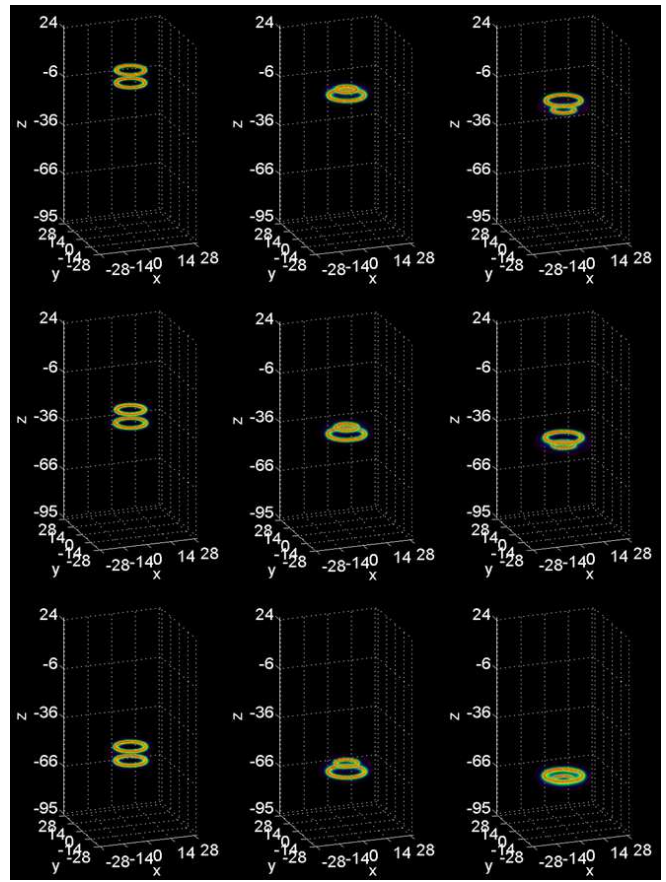}
\caption{
(Color online)
Example of two vortex rings undergoing leapfrogging.
The rings had an initial radius of $d=1$ and were
initially separated in the $z$-direction by a distance $Q=5$.
Times correspond to $t=0,16,27,44,58,68,85,101,110$
(left-to-right, top-to-bottom).
The parameters are the same as in Fig.~\ref{f:headoncol} with
$h=2/3$ and $k=0.045$.}
\label{f:VRlooploop}
\end{figure}

\section{Leapfrogging Vortex Rings}
\label{s:leapfrog}
Let us now consider the interaction between two same sign unit-charge vortex rings ($m_1=m_2=1$) aligned along the $z$ axis.  In classical fluids, the analogous setup exhibits leapfrogging behavior \cite{Konstantinov94,Shashikanth03}.
In Fig.~\ref{f:VRlooploop}, we show an example of two equal-charge vortex rings of radius $d=10$ and initially separated by a distance of $Q=Z_2(0)-Z_1(0)=5$ in the $z$-direction undergoing leapfrogging dynamics.
In the course of our numerical investigation we have found that when $Q$ 
is small (around two or three times the vortex ring's core radius),
the resulting dynamics can be quite different.  For example, instead of 
leapfrogging, the rings can form a bound pair and travel together maintaining
their radii and common speed.
While such dynamical features are intriguing in their own right, for the scope of the current work, we only use results from vortex rings separated with large enough values of $Q$ to result in leapfrogging dynamics.

In order to formulate a reduced model of the leapfrogging interaction of the rings, we start with reviewing the situation in the classical fluid case.  
There, a vorticity field $\boldsymbol\omega$ will induce a velocity field according to
the Biot-Savart law \cite{Batchelor67, Saffman95}
\begin{equation}
\label{vfield1}
\textbf{u}_v(\textbf{x})=\frac{1}{4\pi}\int{\frac{\boldsymbol\omega(\textbf{x}^\prime)\times(\textbf{x}-\textbf{x}^\prime)}{\lvert\textbf{x}-\textbf{x}^\prime\rvert^3}d\textbf{x}^\prime},
\end{equation}
where the integral is computed over all of space.  When considering a vorticity distribution of infinite strength but having a constant circulation $\kappa$ along a closed curve $\textbf{R}$ parametrized by $\ell$,
Eq.~(\ref{vfield1}) becomes
\begin{equation}
\textbf{u}_v(\textbf{x})=\frac{\kappa}{4\pi}\oint{\frac{\textbf{s}\times(\textbf{x}-\textbf{R}(\ell))}{\lvert\textbf{x}-\textbf{R}(\ell)\rvert^3}\,d\ell}
\label{biotsavart}
\end{equation}
where $\textbf{s}$ is the unit tangent vector.  Equation~(\ref{biotsavart}) can, in principle, be evaluated to find the effect of a vortex ring on another vortex ring.  However, when trying to find a ring's self-induced velocity, the integral diverges.  
To circumvent this singularity for the self-induced term the 
localized-induction approximation (LIA)~\cite{DaRios-1906,Betchov-JFM-65,Arms65,Siggia-PhysFluids-85}
is employed by ignoring long-distance effects, yielding the expression
for the self-interacting term:
\begin{equation}
\textbf{u}_v=\frac{(\partial\textbf{R}/\partial \ell)\times(\partial^2\textbf{R}/\partial \ell^2)}{\lvert(\partial\textbf{R}/\partial \ell)\rvert^3}.
\label{localinduction}
\end{equation}
Ultimately, using Eq.~(\ref{biotsavart}) through a similar LIA
for pairwise interactions
and (\ref{localinduction}) for self-interactions,
the following reduced
ordinary differential equations (ODEs) for the effective motion of a
collection of $N$ co-axial vortex rings in a classical fluid 
are obtained~\cite{Konstantinov94,Shashikanth03}:
\begin{align}
\dot Z_i &= \frac{\kappa_i}{4\pi R_i}\left(\ln\frac{8R_i}{r_{c,i}}-C\right)+\frac{1}{\kappa_i R_i}\frac{\partial U}{\partial R_i}\label{vrvelz},\\
\dot R_i &= -\frac{1}{\kappa_i R_i}\frac{\partial U}{\partial Z_i},
\label{vrvelr}
\end{align}
where
$$
U=\frac{1}{2\pi}\sum\limits_{i=1}^N\,\sum\limits_{j>i}^N \kappa_i \kappa_j I_{ij},
$$
and
$$
I_{ij}=\int\limits_0^\pi \frac{R_i R_j \cos \theta\, d\theta}{\sqrt{\left(Z_i-Z_j\right)^2+R_i^2+R_j^2-2R_i R_j \cos\theta}},
$$
$R_i$ and $Z_i$ are the radius and $z$-coordinate of the $i$-th ring, $\kappa_i$ is its circulation or ``vortex strength'' which, in our case, is a constant for each ring, $r_{c,i}$ is its core radius which is time-dependent and follows the relation $r_{c,i}^2(t)R_i(t)=\mbox{constant}$, and $C$ is a constant, which for the case of a classical fluid is given as $C=1/4$
(see Fig.~\ref{f:VR_Drawing}(b)).

We note that the terms involving $U$ relate to the interaction of the rings, while  the first term of Eq.~(\ref{vrvelz}) represents the innate translational velocity of each ring.  If one compares this velocity with that of the quantum vortex ring velocity of Eq.~(\ref{vrvel}), we see that they are quite similar.  We suggest that it is possible to use the results of Eqs.~(\ref{vrvelz}) and (\ref{vrvelr}) for the vortex rings in the NLSE by simply substituting the quantum fluid values such that $C=L_0(m)-1$, $\kappa=4\pi a m$, and $r_c$ from Eq.~(\ref{vrvel}) into Eqs.~(\ref{vrvelz}) and (\ref{vrvelr}).  We point out, that in contrast to the classical fluid leapfrogging picture, in a quantum fluid, the healing length being constant forces $r_c$ to be, approximately, time-independent.
The conservation of the core width during vortex ring evolution
will be briefly addressed below.

\begin{figure}[htb]
\centering
\includegraphics[width=7.5cm]{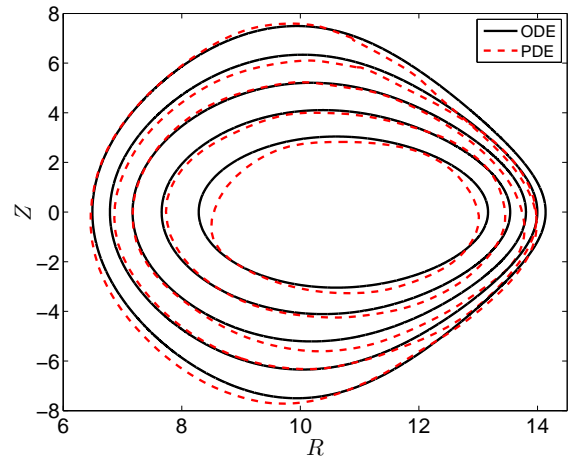}
\caption{
(Color online)
Leapfrogging orbits for two vortex rings in the co-moving $(R,Z)$ plane.
The (black) solid curves depict the results from the
effective, reduced, equations of motion~(\ref{vrvelz}) and (\ref{vrvelr}),
while the (red) dashed curves depict the corresponding results from
those obtained from full 3D integrations of Eq.~(\ref{nlse}).
The rings are initialized with radii $d=11$ and initial separation
distances of $6$, $8$, $10$, $12$, and $14$ (from inner to
outer curves).
Only one period of the leapfrogging motion is shown in a 
reference frame moving at the average velocity of the vortex ring pair
(see Fig.~\ref{f:leapfrog_v}), i.e., a center-of-mass frame.
}
\label{f:rzorbits}
\end{figure}

In Figure~\ref{f:rzorbits}, we show a series of $(R, Z)$ orbits ---in the
center-of-mass frame--- as predicted by Eqs.~(\ref{vrvelz}) and (\ref{vrvelr}) for different initial separation distances along with those obtained through tracking the vortex rings in the integration of the full 3D NLSE system (\ref{nlse}).
As it is clear from the figure, the $(R,Z)$ orbits from the reduced ODE
system are in very good agreement with the ones obtained from the full PDE.
These results stress the usefulness of the reduced system towards capturing the
interaction dynamics of co-axial vortex rings.
We note that while the ODE trajectories are periodic to numerical accuracy,
the PDE results are only approximately periodic (only one ``period''
is shown here).
The deviation from periodicity, in the co-moving reference frame,
in the full PDE numerics might be attributed to several factors
including: internal mode (Kelvin) excitations and
stability of the actual leapfrogging orbits~\cite{Barenghi:arXiv14}.

Figure~\ref{f:rzorbits} suggests the existence of  a stable, co-moving,
equilibrium at the center of the periodic orbits for $Z=0$
and $R$ being constant.
This point corresponds precisely to two overlapping ($Z_1=Z_2$) vortex rings 
of the same radius, namely a vortex ring of charge two. 
In fact the linearization about this co-moving equilibrium
would yield an approximation for the period of the vortex
ring leapfrogging in the limit of small initial separations.
However, as vortex rings
of higher charge are unstable and break up into two unit charge
vortex rings, we will not study them further here
---with the exception of using the theoretically predicted
velocity of the doubly-charged vortex ring as an approximation
for the average leapfrogging velocity.

\begin{figure}[htb]
\centering
\includegraphics[width=\linewidth]{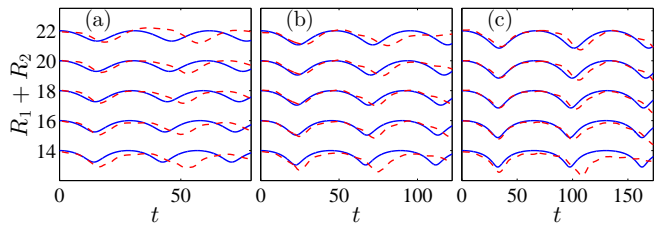}
\caption{
(Color online)
Time dependence for the sum of vortex radii for a pair
of leapfrogging rings.
The (blue) solid curves depict the results from the
effective, reduced, equations of motion~(\ref{vrvelz}) and (\ref{vrvelr}),
while the (red) dashed curves depict the corresponding results from
those obtained from full 3D integrations of Eq.~(\ref{nlse}).
The different panels correspond to vortex rings initially separated by 
(a) $Q=Z_2(0)-Z_1(0)=7$, (b) $Q=Z_2(0)-Z_1(0)=9$, and (c) $Q=Z_2(0)-Z_1(0)=11$ 
units away starting with equal radii $R_1(0)=R_2(0)=11,10,9,8,7$ 
(respective curves from top to bottom).
\label{f:r1r2}
}
\end{figure}

It is interesting to note that the vortex core size in the
NLSE (\ref{nlse}) should be approximately constant provided
the vortex core is not significantly perturbed.
In fact, if the vortex core size is invariant during 
evolution then the total length of the vorticity tube should
be conserved due to conservation of angular momentum.
Therefore, by measuring the total vortex tube length, it is
possible to monitor the degree to
 which the vortex core
width deviates from its unperturbed value.
In the case of leapfrogging vortex rings, where one vortex
is forced inside the other ring, there is a strong mutual
perturbation that can affect the vortex core width.
In Fig.~\ref{f:r1r2} we depict the sum of the vortex radii
for different case examples of
two vortex rings undergoing leapfrogging for both
the effective ODE system and the original PDE. The sum
$R_1+R_2$ is proportional to the total vortex tube
length $L=2\pi(R_1+R_2)$.
As it is clear from the figure, the total vortex length
is not conserved but it oscillates around a mean value
with a relatively small amplitude.
For the effective ODE model, the total vortex tube length
varies between 4\% and 12\%, while for the full 3D model
it varies between 3\% and 8\% for the cases depicted in
Fig.~\ref{f:r1r2}.
This result is a clear indication that the vortex core is
indeed perturbed by the strong mutual interaction
between vortex rings during the leapfrogging dynamics.

It is also worth mentioning that the results in Figs.~\ref{f:rzorbits}
and \ref{f:r1r2} suggest that although the reduced ODEs are
able to capture very well the shape of the leapfrogging
orbits in the $(R,Z)$ plane, there appears to be a systematic
offset (between the ODEs of Eqs.~(\ref{vrvelz})-(\ref{vrvelr})
and the full 3D NLSE) as regards the period for these orbits.
Figure \ref{f:r1r2} shows that for small initial vortex ring
separations [see panel (a)] the offset in the period is larger 
than for large initial vortex ring separations [see panel (c)].
This can be qualitatively understood from the fact that an
initial small separation between vortices induces stronger
interactions between vortex rings bringing them closer to
each other and thus strongly perturbing the vortex core
which was assumed to have zero width in the Biot-Savart law
yielding the ODE approximation.

\begin{figure}[htb]
\centering
\includegraphics[width=7cm]{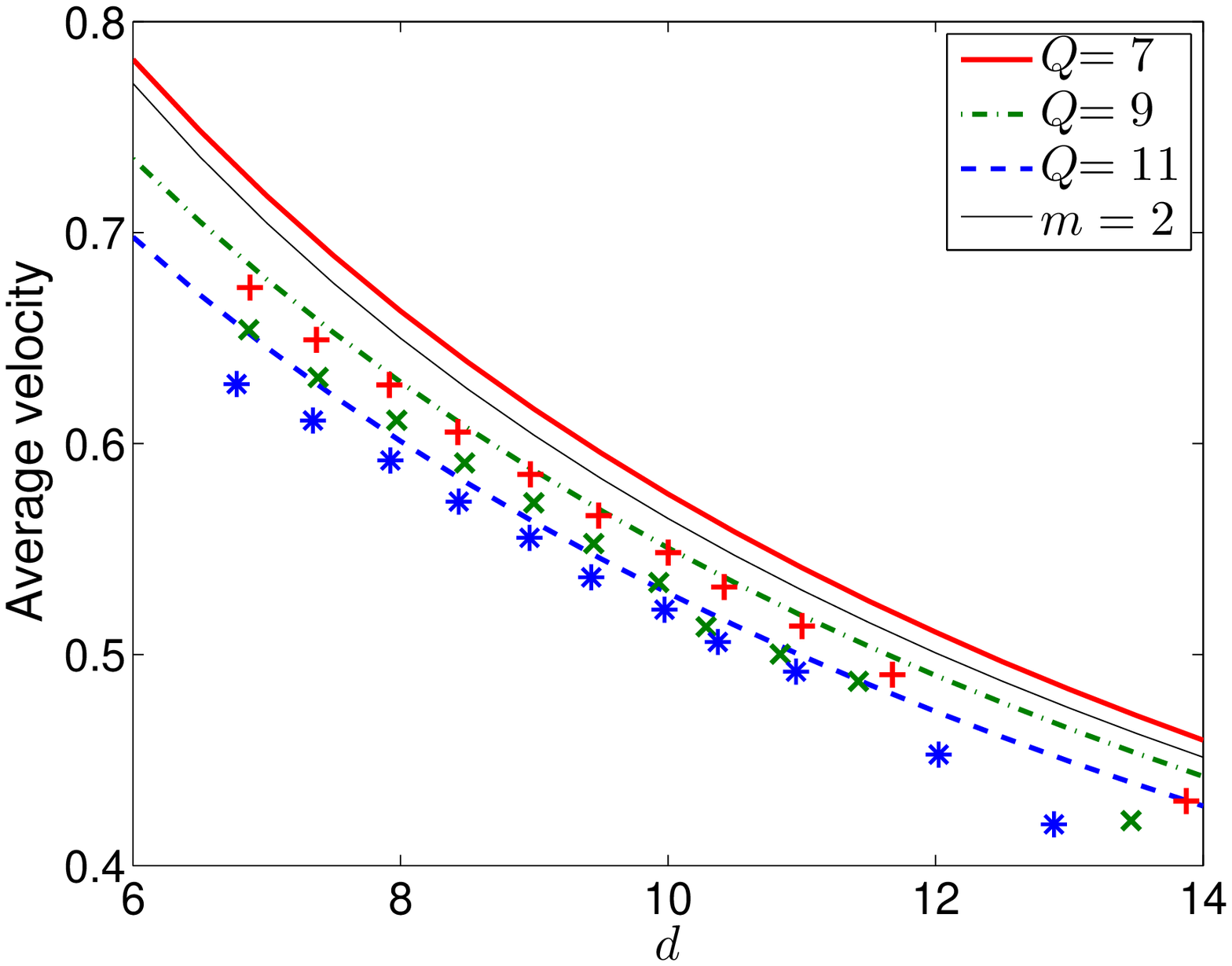}
\includegraphics[width=7cm]{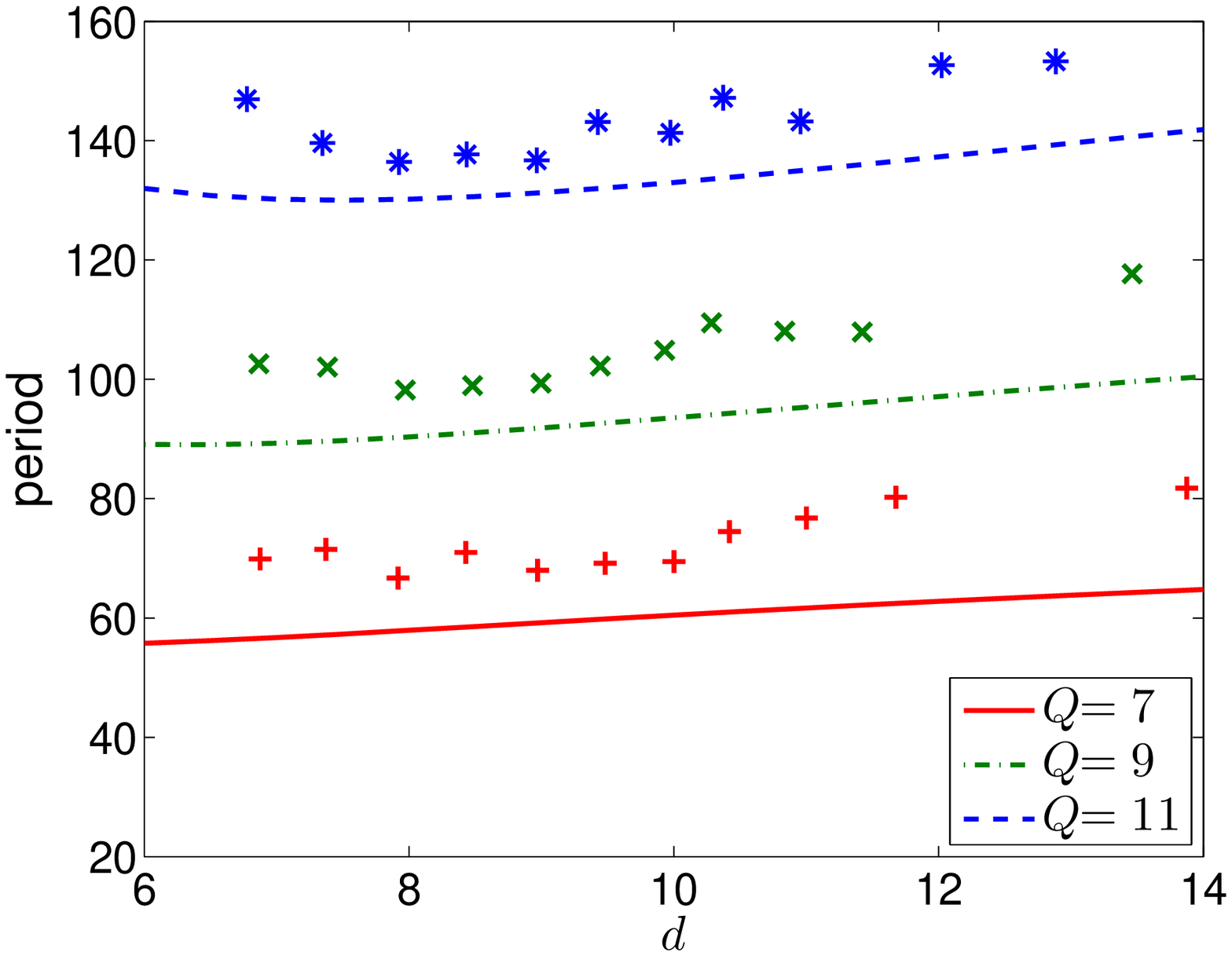}
\caption{
(Color online)
Average velocity (top panel) and looping period  (bottom panel)
of leapfrogging vortex rings as a function of the
ring radii ($d$) for several separation distances ($Q$).
The lines correspond to the results obtained through numerical
integration of the effective equations of motion~(\ref{vrvelz})
and (\ref{vrvelr}) and the symbols correspond to the full
numerical integration of the NLSE model (\ref{nlse}).
For comparison we also include (thin black line)
the velocity of a vortex ring of
charge $m=2$ using Eq.~(\ref{vrvel}).
The horizontal axis corresponds to the initial radius of the pair of
identical vortex rings, while the different colors
for the thick lines correspond
(top to bottom) to initial separation distances of 7 (red solid line
and + symbols), 9 (green dotted line and $\times$ symbols),
and 11 (blue dashed line and * symbols).}
\label{f:leapfrog_v}
\end{figure}

Two relevant quantities of the leapfrogging rings that we can use to
further validate the reduced model are the average ring-pair $z$-velocity
and the rings' period of oscillation.  We measure the average ring-pair
$z$-velocity by tracking the $z$ value of the center position of a
$(y,z)$ cut of the rings over time and then perform a least-squares
linear fit.  This is done for both the full NLSE simulations, and for
the ODE integration of the reduced system.  In Fig.~\ref{f:leapfrog_v},
the results are presented as a function of $d$ for a few values of the
initial $z$ separation distance, $Q$.
We observe that the average velocity of the leapfrogging rings obtained
in the PDE model is always lower than that obtained in the effective ODE
model, the percent difference ranges between $1.6\%$ and $7.2\%$.
It is interesting to compare the average velocity of two leapfrogging
rings with the velocity predicted for a vortex ring of charge $m=2$
from Eq.~(\ref{vrvel}). As depicted by the black thin line in the top
panel of Fig.~\ref{f:leapfrog_v}, the velocity of a vortex ring of
charge $m=2$ (evaluated at the initial radii for
the leapfrogging vortex ring pair)
closely resembles the average velocity of a vortex ring
pair under leapfrogging evolution. This stems from the fact that,
in the far field, a vortex ring pair can be approximated by a
vortex ring of charge $m=2$. In fact, after closer inspection
of the results in the top panels of Fig.~\ref{f:leapfrog_v}, the
velocity of the doubly-charged vortex ring is closer to the
velocity for the vortex pairs when their separation is small, i.e.,
when they are closer to each other and thus closer to a doubly-charged
vortex ring.
While,  this relation could be further explored by taking
the average radius along one leapfrogging period instead of the
initial radii, the main phenomenology does not change significantly.

In the bottom panel of Fig.~\ref{f:leapfrog_v} we show the results of
measuring the period of oscillation of the leapfrogging rings from both the full NLSE simulations and the ODE reduced model for the same ring separations and radii.
We see that the period of oscillation for the PDE model
is always longer than that of the ODE model, and the percent difference
in period ranges from $4.2\%$ to $28.5\%$ percent, which, while clearly worse
than the velocity comparison, is not unreasonable.
It is interesting to note that while the period of the leapfrogging rings changes depending on the $z$-distance between the rings, it remains
fairly constant for vortex ring pairs of {\em different} radii at the same separation distance.

From the velocity and period results, we see that there are obviously 
some differences between the quantum and classical fluid regimes including 
the modified values that result in lower average pair velocities and longer periods of oscillation than those
predicted by our current reduced equations of motion.  These are worthy
of further investigation, in order to improve the accuracy of the reduced 
equations of motion.
Nevertheless, the trends of the two sets of results are very similar, and 
their magnitudes
are comparable, which suggests that our effective description provides a 
valuable means of examining the numerical (and, presumably, also the
experimental) data and, as such, is useful in its own right as well as
worthy of further investigation.

Finally, it is worth pointing out that there has been recent efforts
in describing the leapfrogging dynamics of vortex ring
bundles in classical fluids using the vortex filament
method~\cite{Barenghi:arXiv14}.
In particular, it is found that the leapfrogging dynamics
is weakly unstable for small number of vortex rings.
In our numerics we have found similar instabilities, that
depending on the separation and radii of the vortex rings,
may not be visible until tens of leapfrogging periods.
Therefore, although weakly unstable, the time required for
the instability to manifest itself might be long
enough to be able to observe leapfrogging during the
typical lifetime of current BEC experiments.
%

It is also important to mention that the 2D equivalent 
of leapfrogging has been studied in the case of
point vortices \cite{Acheson-EJP-00,Aref-PF-13}.
This 2D leapfrogging corresponds to two vortex dipoles 
(vortex and anti-vortex pairs) and it is found to display
three distinct regions depending on the distance between
pairs (relative to the internal distance between
vortices in each pair) as follows. 
(a) The small separation region does not yield leapfrogging.
(b) Intermediate separations yield {\em unstable} leapfrogging.
(c) Large separation distances yield {\em stable} leapfrogging.
The 2D vortex dipole leapfrogging is strongly connected
to 3D vortex ring leapfrogging since any 2D cut passing through the
axes of the vortex rings will produce a 2D pair of vortex dipoles.
Therefore, 2D vortex dipole dynamics is the basic
ingredient responsible for vortex ring leapfrogging.
Therefore, it would be interesting to test if the
vortex ring leapfrogging also displays the three regions described above
and whether the weak instability that we have observed
can in fact be eliminated by using larger vortex ring radii.
This avenue of research is currently being examined and a systematic
investigation
will be reported in a future publication.

\section{Conclusions}
\label{s:conclu}
We studied the dynamics and interactions of
vortex rings in 
the nonlinear Schr\"odinger equation as a model for a 
homogeneous Bose-Einstein condensate.
Upon corroborating the well-known single vortex ring evolution in 
our 3D NLSE framework,
we focused on two dynamical scenarios: scattering and
leapfrogging.
For the former, we considered the scattering
collision of two co-planar, opposite charge, vortex rings.
Upon their  collision, these scatter at an
angle depending on the initial conditions.
We found a surprisingly simple phenomenological rule for the
scattering angle as a function of the vortex ring radii and the
initial distance (impact parameter) between the vortex rings.
For the second scenario, we followed the dynamics of a
pair of co-axial vortex rings of the same charge.
We proposed a modification of the effective equations
of motion for the vortex ring interactions in a
classical superfluid. The modification was based on correcting the
self-induced velocity using previous results for NLSE settings
in a quantum superfluid.
We compared the resulting effective equations of motion
against full 3D simulations of the original NLSE model for
several cases of leapfrogging evolution and found good
agreement between the two.
Finally, we found a monotonic decrease for the average velocity of the
leapfrogging as the radii of the ring increase and/or as their
initial separation increases.
We also found that the leapfrogging period remains approximately
constant for vortex ring pairs of different radii starting
at the same separation distance, while the period increases
with initial separation distance for fixed radii.

Future work will be directed
towards refining the reduced equations of motion to better account
for the differences between classical and quantum fluids
(especially in the context of the term encoding the interaction
between the different rings), and
expanding them to describe more general configurations of vortex
rings, as opposed to only those which are co-axial. 
It would also be interesting to understand and quantify
the instability (and its range of relevance over initial
conditions and system parameters) observed in the leapfrogging dynamics,
not only for a pair of vortex rings but also, for vortex ring
bundles~\cite{Barenghi:arXiv14}.
Another avenue for future exploration consists of using
classical fluid-inspired effective equations of motion
(Biot-Savart law) to obtain reduced ODEs describing the
effects of slow-down created by internal excitation (Kelvin)
modes~\cite{maggioni10,Helm11,Helm10} and, in more general
terms, the effects of these modes on the interactions and
scattering between vortex rings.
We are currently exploring these directions and will report
on them in future works.

\section*{Appendix: Computation of the vortex core parameter}
The ``vortex core parameter'' $L_0(m)$ is defined as the convergent part of the energy per unit length of a vortex line.
In this appendix, we simply state the equations used to compute $L_0(m)$.  For full details of the derivation of the equations, we refer the reader to Refs.~\onlinecite{Helm10,RMC-DISS}.

For the NLSE in the form of Eq.~(\ref{nlse}), the vortex core parameter is found by numerically integrating
\begin{alignat}{2}
\label{L0}
L_0(m) &= \ln(r_c)+\frac{s}{am^2\Omega}\left(a \int_0^{\infty}\!\left(\frac{df}{dr}\right)^2 r\,dr  \right.
\\
&-\frac{s}{2}\int_0^\infty\! \left(\frac{\Omega}{s}-f^2(r)\right)^2 r\,dr \notag
\\
&-\left.2am^2\int_0^\infty\! f(r)\frac{df}{dr}\ln(r)\,dr\right),
\notag
\end{alignat}
where $f(r)$ is found numerically by solving the ODE
$$
-\left(\Omega + \frac{am^2}{r^2}\right)f(r) + a\left(\frac{1}{r}\frac{df}{dr} + \frac{d^2f}{dr^2}\right) + sf^3(r)=0,
$$
and where we note that for a given $m$, $L_0(m)$ is invariant over the parameters $a$, $s$, and $\Omega$.

\begin{table}[htb]
\centering
\begin{tabular}{|c|c||c|c|} \hline
$m$  & $L_0(m)$     & $m$  & $L_0(m)$  \\ \hline
$1$  & ~$0.380868$~   & $6$  & ~$0.022954$~\\
$2$  & $0.133837$   & $7$  & $0.017785$\\
$3$  & $0.070755$   & $8$  & $0.014247$\\
$4$  & $0.044567$   & $9$  & $0.011713$\\
$5$  & $0.030981$   &$10$  & $0.009833$\\
\hline
\end{tabular}
\caption{
Values of the vortex core parameter $L_0$ for charges $m\in[1,10]$.
The values are computed through numerical integration of Eq.~(\ref{L0}).}
\label{t:l0ofm}
\end{table}

\begin{figure}[htb]
\centering
\includegraphics[width=7.5cm]{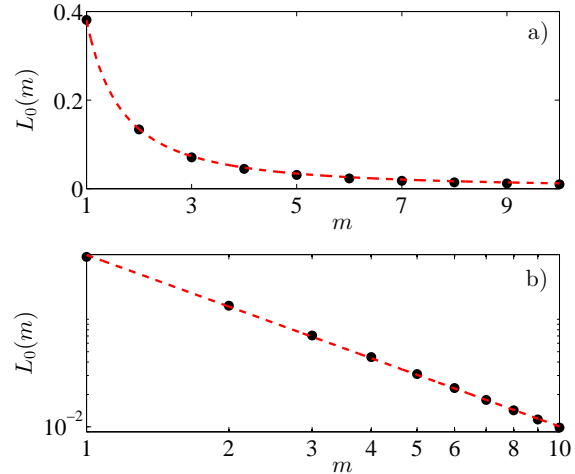}
\caption{
(Color online)
Computed vortex core parameters for charges $m=1$ to $m=10$.  The dots are the 
results of integrating Eq.~(\ref{L0}). 
a) The dashed line is the fitted curve of Eq.~(\ref{l0ofm}).
b) Same as panel a) but in log-log plot and using the best linear
fit to the log-log of data giving the power law $L_0 \propto m^{-1.5956}$.
\label{f:l0}
}
\end{figure}

We compute the integrals in Eq.~(\ref{L0}) using a simple trapezoid integration on a numerically-exact $f(r)$ computed on a domain of $r=[0,500]$, with a spatial step size of $\Delta r=0.0075$ and NLSE parameters $a=1$, $s=-1$, and $\Omega=-1$.
The values of $L_0$ for charges $m=1,...,10$ are given in Table.~\ref{t:l0ofm}.
We note that the values of $L_0$ for $|m|>3$ have not been previously reported, and those for $m=2$ and $m=3$ are given with more accuracy than reported in Ref.~\onlinecite{VLINE-L058}.

It is intriguing to note that one can produce a very close power-law fit yielding an 
approximation of $L_0(m)$ given by
\begin{equation}
\label{l0ofm}
L_0(m) \approx L_0(1)\,m^{-3/2},
\end{equation}
which may indicate the possibility that a closed-form expression for $L_0(m)$ is achievable.  
In Fig.~\ref{f:l0} we show a comparison of the values of $L_0$ given in Table~\ref{t:l0ofm} 
to those given by Eq.~(\ref{l0ofm}).

\acknowledgments{
We gratefully acknowledge the support of NSF-DMS-0806762
and NSF-DMS-1312856, NSF-CMMI-1000337, as well as from 
the AFOSR under grant FA950-12-1-0332, the
Binational Science Foundation under grant 2010239, from the
Alexander von Humboldt Foundation and the FP7, Marie 
Curie Actions, People, International Research Staff 
Exchange Scheme (IRSES-606096). 
RMC and JDT acknowledge the support of the Computational Science
Research Center (CSRC) at SDSU.
}

\end{document}